
\input harvmac
\noblackbox
\def\cR{{\cal R}}
\def\C{{\bf C}}
\def\O{{\bf O}}
\def\I{{\bf I}}

\def\N{{\bf N}}
\def\B{{\bf B}}

\lref\gibmae{
G.W. Gibbons and K. Maeda,
``Black Holes and Membranes in Higher Dimensional Theories with
Dilaton Fields,''
{\it Nucl. Phys.} {\bf B298} (1988) 741.}
\lref\ghs{
D. Garfinkle, G.T. Horowitz, and A. Strominger,
``Charged Black Holes in String Theory,''
{\it Phys. Rev.} {\bf D43} (1991) 3140; {\bf D45} (1992) 3888(E).}
\lref\stw{
A. Shapere, S. Trivedi, and F. Wilczek,
``Dual Dilaton Dyons,''
{\it Mod. Phys. Lett.} {\bf A6} (1991) 2677;
A.G. Agnese and M. La Camera,
``General Spherically Symmetric Solutions in Charged Dilaton Gravity,''
{\it Phys. Rev.} {\bf D49} (1994) 2126.
}
\lref\grehar{
R. Gregory and J.A. Harvey,
``Black Holes with a Massive Dilaton,''
{\it Phys. Rev.} {\bf D47} (1993) 2411, hep-th/9209070.}
\lref\horhora{
J.H. Horne and G.T. Horowitz,
``Black Holes Coupled to a Massive Dilaton,''
{\it Nucl. Phys.} {\bf B399} (1993) 169, hep-th/9210012.}
\lref\polwil{
S.J. Poletti and D.L. Wiltshire,
``The Global Properties of Static Spherically Symmetric Charged Dilaton
Space-times with a Liouville Potential,''
{\it Phys. Rev.} {\bf D50} (1994) 7260, gr-qc/9407021.}
\lref\ozetah{
M. Ozer and M.O. Taha,
``Exact Solutions in String Motivated Scalar Field Cosmology,''
{\it Phys. Rev.} {\bf D45} (1992) 997.}
\lref\easther{
R. Easther, ``Exact Superstring Motivated Cosmological Models,''
{\it Class. Quant. Grav.} {\bf 10} (1993) 2203.}
\lref\wald{
R.M. Wald, {\it General Relativity} (University of Chicago Press,
Chicago, 1984).}
\lref\chaman{
K.C.K. Chan and R.B. Mann,
``Static Charged Black Holes in (2+1)-dimensional Dilaton Gravity,''
{\it Phys. Rev.} {\bf D50} (1994) 6385, gc-qc/9404040.}
\lref\broyor{
J.D. Brown and J.W. York,
``Quasilocal Energy and Conserved Charges Derived from the Gravitational
Action,''
{\it Phys. Rev.} {\bf D47} (1993) 1407;
J.D. Brown, J. Creighton, and R.B. Mann,
``Temperature, Energy and Heat Capacity of Asymptotically Anti-de Sitter
Black Holes,''
{\it Phys. Rev.} {\bf D50} (1994) 6394, gc-qc/9405007.}
\lref\visser{
M. Visser,
``Dirty Black Holes: Entropy as a Surface Term,''
{\it Phys. Rev.} {\bf D48} (1993) 5697, hep-th/9307194.}
\lref\mann{
R.B. Mann,
``Conservation Laws and 2-d Black Holes in Dilaton Gravity,''
{\it Phys. Rev.} {\bf D47} (1993) 4438, hep-th/9206044.}
\lref\gidstr{
S.B. Giddings and A. Strominger,
``Dynamics of Extremal Black Holes,''
{\it Phys. Rev.} {\bf D46} (1992) 627, hep-th/9202004;
R. Kallosh and A. Peet,
``Dilaton Black Holes near the Horizon,''
{\it Phys. Rev.} {\bf D46} (1992) 5223, hep-th/9209116;
Y. Peleg,
``4-Dimensional And 2-Dimensional Evaporating Dilatonic Black
Holes,''
{\it Mod. Phys. Lett.} {\bf A9} (1994) 3137.}
\lref\makshia{
T. Maki and K. Shiraishi,
``Multi-black Hole Solutions in Cosmological Einstein-Maxwell-dilaton
Theory,''
{\it Class. Quantum Grav.} {\bf 10} (1993) 2171.}
\lref\kastra{
D. Kastor and J. Traschen,
``Cosmological Multi-black Hole Solutions,''
{\it Phys. Rev.} {\bf D47} (1993) 5370, hep-th/9212035.}
\lref\wilt{
D.L. Wiltshire,
``Spherically Symmetric Solutions in Dimensionally Reduced Space-times with
a Higher Dimensional Cosmological Constant,''
{\it Phys. Rev.} {\bf D44} (1991) 1100.}
\lref\kimcho{
S.W. Kim and B.H. Cho,
``Higher Dimensional Black Hole Solution with Dilaton Field,''
{\it Phys. Rev.} {\bf D40} (1989) 4028.}
\lref\horhorb{
J.H. Horne and G.T. Horowitz,
``Cosmic Censorship and the Dilaton,''
{\it Phys. Rev.} {\bf D48} (1993) 5457, hep-th/9307177.}
\lref\horhorc{
J.H. Horne and G.T. Horowitz,
``Rotating Dilaton Black Holes,''
{\it Phys. Rev.} {\bf D46} (1992) 1340, hep-th/9203083.}
\lref\makshib{
T. Maki and K. Shiraishi,
``Three-dimensional Black Holes and Solitons in Higher Dimensional Theories
with Compactification,''
{\it Class. Quant. Grav.} {\bf 11} (1994) 2781.}
\lref\gershon{
D. Gershon, ``Semiclassical {\it vs.}~Exact Solutions of Charged Black Hole
in Four Dimensions and Exact $O(d,d)$ Duality,'' {\it Nucl. Phys.} {\bf B421}
(1994) 80, hep-th/9311122.}
\lref\bechlech{
O. Bechmann and O. Lechtenfeld, ``Exact Black-Hole Solution With
Self-Interacting
Scalar Field,'' gr-qc/9502011.}
\lref\witten{E. Witten,
``On String Theory and Black Holes,''
{\it Phys. Rev.} {\bf D44} (1991) 314.}
\lref\fzb{V.P. Frolov, A.I. Zelnikov, and U. Bleyer,
``Charged Rotating Black Hole from Five Dimensional Point of View,''
{\it Ann. Phys. (Leipzig)} {\bf 44} (1987) 371.}
\lref\sen{A. Sen,
``Rotating Charged Black Hole Solution in Heterotic String Theory,''
{\it Phys. Rev. Lett.} {\bf 69} (1992) 1006,
hep-th/9204046.
}
\lref\townsend{
G.W. Gibbons, G.T. Horowitz, and P.K. Townsend, ``Higher-Dimensional
Resolution of Dilaton Black Hole Singularities,'' hep-th/9410073.
}

\Title{\vbox{\baselineskip12pt\hbox{DAMTP/R-95/7}
\hbox{WATPHYS TH-95/02}
\hbox{gr-qc/9502042}
}}
{\vbox{
\centerline{Charged Dilaton Black Holes}
\medskip
\centerline{with Unusual Asymptotics}
}}

\centerline{
{Kevin C.K. Chan}\footnote{$^\dagger$}{Email address:
kckchan@astro.uwaterloo.ca},
{James H. Horne}\footnote{$^\ddagger$}{Email address: jhh20@amtp.cam.ac.uk},
and {Robert B. Mann}\footnote{$^*$}{On leave from University of
Waterloo. Email address: rbm20@amtp.cam.ac.uk}}
\vskip .12in
\centerline{\sl $^{\dagger,*}$ Department of Physics}
\centerline{\sl University of Waterloo}
\centerline{\sl Waterloo, Ontario}
\centerline{\sl Canada N2L 3G1}
\vskip.12in
\centerline{\sl $^{*,\ddagger}$ D.A.M.T.P.}
\centerline{\sl	Cambridge University}
\centerline{\sl	Silver Street}
\centerline{\sl	Cambridge CB3 9EW}
\centerline{\sl	Great Britain}

\bigskip
\centerline{\bf Abstract}
We present a new class of black hole solutions in
Einstein-Maxwell-dilaton gravity in $n \ge 4$ dimensions. These
solutions have regular horizons and a singularity only at the
origin. Their asymptotic behavior is neither asymptotically flat nor
(anti-) de Sitter. Similar solutions exist for certain Liouville-type
potentials for the dilaton.

\Date{Feb.~27, 1995}

\newsec{Introduction}

There has been great interest in recent years in $n$-dimensional
($n\geq 4$) dilaton gravity. It is important to investigate how the
properties of black holes are modified when a dilaton is
present. $n$-dimensional static spherically symmetric $(SSS)$
electrically or magnetically charged dilatonic black hole solutions
were first analyzed in some generality by Gibbons and Maeda~\gibmae,
and the four-dimensional versions were rediscovered and clarified in
an elegant note by Garfinkle, Horowitz and Strominger~\ghs.  A number
of authors have considered variations, such as $SSS$ dyonic black hole
solutions~\stw, or black holes when the dilaton acquires a
mass~\refs{\grehar,\horhora}.

In this paper, we obtain and discuss a new class of electrically (or
magnetically) charged $SSS$ black hole solutions to $n$-dimensional
dilaton theories of gravity. The metrics associated with these
solutions are neither asymptotically flat nor asymptotically (anti-)
de Sitter; however the only curvature singularities are at the origin.

We first consider the four-dimensional action in which gravity is
coupled to a dilaton and Maxwell field with an action
\eqn\esdef{
S=\int \! d^4x {\sqrt{-g}} (\cR-2(\nabla\phi)^2-V(\phi)-e^{-2a\phi}F^2) \;,
}
where $\cR$ is the scalar curvature, $F^2=F_{\mu\nu}F^{\mu\nu}$ is the
usual Maxwell contribution, and $V(\phi)$ is a potential for
$\phi$. The constant $a$ governs the coupling of $\phi$ to
$F_{\mu\nu}$. We will consider three special cases: (i) $V(\phi)=0$,
(ii) $V(\phi)={2\Lambda}e^{2b\phi}$ and (iii) $V(\phi)=2\Lambda_1
e^{2b_1\phi} +2\Lambda_2 e^{2b_2\phi}$. The first case corresponds to
the action considered in~\refs{\gibmae,\ghs}. When $a=1$, it reduces
to the four-dimensional low-energy action obtained from string theory
in terms of Einstein metric. Case (ii) corresponds to a Liouville-type
gravity. We will refer to $\Lambda$ as the cosmological constant,
although in the presence of a non-trivial dilaton field, the spacetime
does not behave as either de Sitter ($\Lambda>0$) or anti-de Sitter
($\Lambda<0$).  It has recently been shown that with the exception of
a pure cosmological constant potential ({\it{i.e.}}, $b=0$), no
asymptotically flat ($\Lambda=0$), asymptotically de Sitter or
asymptotically anti-de Sitter $SSS$ solutions to the field equations
associated with case (ii) exist~\polwil. However, non-static
generalizations of the Kastor-Traschen-type~\kastra\ cosmological
multi-black hole solutions are known~\makshia. The potential in case
(iii) was previously investigated by a number of
authors~\refs{\ozetah,\easther} in the context of $FRW$ scalar field
cosmologies. This kind of potential function can be obtained when a
higher-dimensional theory is compactified to four-dimensional
spacetime, including various supergravity and string models. In case
(i), the non-asymptotically flat black hole solutions we derive in
this paper were previously obtained by several authors but they failed
to recognize that the solutions admit regular horizons (see
section~3). In cases (ii) and (iii), we will construct the first
examples of $SSS$ black hole solutions.

For the $n$-dimensional cases, we generalize the action \esdef\ to
\eqn\eddef{
S=\int \!d^n x \sqrt{-g}\bigl(\cR-{4\over n-2}(\nabla\phi)^2
- V(\phi)-e^{-{4a\phi\over n-2}}F^2 \bigr) \;.
}
We will first derive the four-dimensional solutions and then
generalize them to $n$ dimensions using~\eddef.  For simplicity, we
will only consider pure electrically (or magnetically) charged cases.

The solutions we derive in this paper are not vacuum solutions in a
strict sense since the action \esdef\ (or \eddef) contains a static
``dilaton fluid'' whose energy momentum tensor is nowhere vanishing.
Here we adopt special dilaton fluid models in which the metric
approaches neither the Minkowski nor (anti-) de Sitter metrics at
spatial infinity. Although the dilaton field diverges for large $r$
($\phi\propto \log r$ where $r$ is the usual radial coordinate), the
quasilocal mass is finite for all values of $r$ outside the event
horizon in every black hole solution.  For all the black hole
solutions, the Ricci and Kretschmann scalars are everywhere finite
except at $r=0$ but those singularities are hidden by event
horizons. It can also be checked that all terms in the action \esdef\
(with the any of the potentials (i)-(iii)) are finite for
$\infty>r\geq r_h$ (where $r_h$ denotes the event horizon), and vanish
as $r\rightarrow\infty$ in all the black hole solutions we obtain.  As
a consequence we believe that our black hole solutions are of some
physical interest.

The organization of this paper is as follows. In section~2, we adopt a
$SSS$ ansatz and then write down the equations of motion for the
action~\esdef.  The formula for calculating the quasilocal mass of a
$SSS$ black hole metric is briefly reviewed.  We first consider a
vanishing potential $V$ in section~3.  When $a=1$, the solution
corresponds to a non-asymptotically flat string black hole in terms of
the Einstein metric.  We will discuss the solution in terms of the
string metric as well.  Regardless of whether $a=1$ or not, we will
see that these black hole metrics have no inner horizon. Furthermore,
there is no extremal limit for the electric charge. More precisely,
the charge can take any finite value without causing the event horion
to either vanish or become singular. Such a property is not shared by
the asymptotically flat black holes in~\refs{\gibmae,\ghs} or the
Reissner-Nordstrom solution in General Relativity. In section~4 we
solve the field equations for a simple Liouville-type potential and
obtain three families of electrically charged Liouville black hole
solutions.  In section~5, we consider the double Liouville potential
and construct a family of black holes with two cosmological couplings
for general $b_1$ and $b_2$.  Finally, we solve the field equations
for the action \eddef\ and generalize the above four-dimensional
solutions to $n$ dimensions in section~6.  We summarize our results
and discuss possible future work in the concluding section.

Our conventions are as in Wald~\wald.

\newsec{Field Equations}

We first consider the four-dimensional action \esdef.  Varying \esdef\
with respect to the metric, Maxwell, and dilaton fields, respectively,
yields (after some manipulation)
\eqna\eeom
$$\eqalignno{
\cR_{ab} = & 2 \partial_a \phi \partial_b \phi +
{1 \over 2} g_{ab} V  + 2 e^{ -{2 a \phi}} \left( F_{ac}
{F_b}^c - {1 \over 4} g_{ab} F_{cd} F^{cd} \right) \;, &
\eeom{a} \cr
0 = & \partial_a \left[ \sqrt{-g} e^{- {2 a \phi}} F^{ab}
\right] \;, & \eeom{b} \cr
\nabla^2\phi = & {1 \over 4} {\partial V \over \partial
\phi} - {a \over 2} e^{-{2a \phi}} F_{ab} F^{ab} \; . &
\eeom{c} \cr
}$$
We wish to find $SSS$ solutions to \eeom{} which admit non-singular
horizons.  In $3+1$ dimensions, the most general such metric has two
degrees of freedom, and can be written in the form
\eqn\emeta{
ds^{2}=-U(r)dt^2+{dr^2\over U(r)}+R^2(r)d\Omega^2 \;.
}
For most of this paper, we will specialize to Maxwell fields
corresponding to an isolated electric charge. The Maxwell
equation~\eeom{b} can be integrated to give
\eqn\efsol{
F_{tr}={Qe^{2a\phi}\over R^2} \;,
}
where $Q$ is the electric charge, defined through the integral
$-{1\over 8\pi}
{\int}_{S}e^{-2a\phi}\epsilon_{\mu\nu\alpha\gamma}F^{\alpha\gamma}$,
where $S$ is any two-sphere defined at spatial infinity and
${\bf{\epsilon}}$ is the volume element (see~\wald).  With the
metric~\emeta\ and Maxwell field~\efsol, the equations of
motion~\eeom{} reduce to three independent equations:
\eqna\eeomb
$$\eqalignno{
{1 \over R^{2}} {d \over dr}\left[ R^{2} U {d \phi \over dr}
\right] = & {1 \over 4} {d V \over d\phi} + a e^{{2 a \phi}}
{Q^2 \over R^{4}} \; , & \eeomb{a} \cr
{1 \over R} {d^2 R \over dr^2} = & - \left( {d
\phi \over dr} \right)^2 \; , & \eeomb{b} \cr
{1 \over R^{2}} {d \over dr} \left[ U {d \over dr} \left( R^{2}
\right)
 \right] = & 2 {1 \over R^2} - V - 2 e^{{2 a
\phi}} {Q^2 \over R^{4}} \; . & \eeomb{c} \cr
}
$$
If we had instead chosen the Maxwell field to be an isolated magnetic
charge at the origin, $F_{\theta\phi} = Q \sin \theta$, then the
equations of motion~\eeomb{} would be unchanged except for the
replacement $a \rightarrow -a$. Thus, for any electrically charged
solution we obtain for arbitrary $a$, the corresponding magnetically
charged solution can be found by simply replacing $a \rightarrow -a$
everywhere (this symmetry is not valid for $n > 4$). Thus we will only
specifically mention the magnetically charged solution in one
instance.

It is easy to see in~\eeomb{b} that if $R(r) = r$, then $\phi$ must be
constant.  Thus one generally cannot have both $g_{tt}=-1/g_{rr}$ and
the angular part of the metric of the form $r^2d\Omega^2$ when there
is a non-trivial dilaton present. However in~\chaman, a variety of
solutions to the 2+1 dimensional action analogous to \esdef\ were
obtained by making the ansatz
\eqn\ehansatz{
R(r) =\gamma r^{N} \; ,
}
where $\gamma$ and $N$ are constants. Using~\ehansatz\ with \eeomb{b}
immediately forces $\phi$ to have the form
\eqn\ephiansatz{
\phi(r) = \phi_0 + \phi_1 \log r \; .
}
A dilaton of the form~\ephiansatz\ is common in two and three
dimensional solutions, but is generally regarded as a sign of
pathology in four or higher dimensions. We present in this paper
solutions in four and higher dimensions using~\ehansatz\ and
\ephiansatz\ that have finite mass, curvature, and charge at large
$r$, and thus should be regarded as physically interesting.

Because the metrics presented in this paper will not be asymptotically
flat, we must use the quasilocal formalism to define the mass of the
solutions (for a detailed review of the concepts of quasilocal mass,
see~\broyor). If a $SSS$ metric is written in the form
\eqn\emetb{
ds^2=-W^2(r)dt^2+{dr^2\over V^2(r)}+r^2d\Omega^2_{n -2} \;,
}
and the matter action contains no derivatives of the metric, then the
quasilocal mass is given by
\eqn\emquas{
{\cal M} = { n - 2 \over 2} r^{n - 3} W(r) (V_0(r) - V(r))\;.
}
Here $V_0(r)$ is an arbitrary function which determines the zero of
the energy for a background spacetime and $r$ is the radius of the
spacelike hypersurface boundary. When the spacetime is asymptotically
flat, the $ADM$ mass $M$ is the ${\cal M}$ determined in \emquas\ in
the limit $r\rightarrow\infty$. If no cosmological horizon is present,
the large $r$ limit of~\emquas\ is used to determine the mass.  If a
cosmological horizon is present once cannot take the large $r$ limit
to identify the quasilocal mass. However, one can still identify the
``small mass'' parameter in the solution~\broyor.

\newsec{Solutions with $V(\phi)=0$}

We begin by looking for four-dimensional solutions with $V(\phi) = 0$.
The system of equations~\eeomb{} has been completely solved
before~\refs{\gibmae,\horhora,\polwil}, but always with the aim of
finding asymptotically flat solutions. The solutions we will present
in this section were implicitly present in those works, but have not
before been written down explicitly.

\subsec{The string case, $a=1$}

We first consider the string theoretic case, where $a=1$ in \esdef.
This case has a number of interesting special features.
Using~\ehansatz\ in \eeomb{}, we find the solution
\eqna\estring
$$\eqalignno{
R(r) = & \gamma r^{1 \over 2} \; , & \estring{a} \cr
U(r) = & {1 \over \gamma^2} (r - 4 M) \; , & \estring{b} \cr
\phi(r) = & - {1 \over 2} \log\left( {2 Q^2 \over \gamma^2} \right) +
{1 \over 2} \log r \; . & \estring{c}
}$$
By performing a transformation $\gamma r^{1 \over 2} \rightarrow r$,
we can write the metric
\eqn\emetb{
ds^2 = - {1 \over \gamma^4} (r^2 - 4 \gamma^2 M) dt^2 + {4 r^2 \over
r^2 - 4 \gamma^2 M} dr^2 + r^2 d\Omega^2 \; ,
}
with a dilaton
\eqn\ephib{
\phi(r) = - {1 \over 2} \log (2 Q^2) + \log r \; .
}
$M$ is the quasilocal mass calculated from \emquas\ and~\emetb. A
magnetically charged version of~\emetb\ for a specific value of $M$
and $Q$ can be found in~\gershon.

The solution~\emetb\ has several interesting properties.  First, it is
easy to show that $\cR = { r^2 - 4 \gamma^2 M \over 2 r^4}$. Thus,
\emetb\ is well behaved at the event horizon $r_h = 2 \gamma
\sqrt{M}$, and has a singularity at $r=0$. Even though $\phi \propto
\log r$ for large $r$, both the Kretschmann scalar and $\cR$ vanish as
$r \rightarrow \infty$. Since the mass, charge, and curvature all
remain finite as $r \rightarrow \infty$, the solution~\emetb\ is
well-behaved there.  Second, \emetb\ has no smooth transition to the
$Q=0$ case because of~\ephib. However, no extremal limit
exists. Finally, it is straightforward to check that the causal
structure of~\emetb\ is exactly the same as the Schwarzschild case.
Thus the singularity is spacelike.

Strings do not couple to the Einstein metric $g_{ab}$ but rather to
the string metric $\tilde{g}_{ab} = e^{2\phi}g_{ab}$, where the
action~\esdef\ reads
\eqn\esact{
S=\int\! d^4x\sqrt{-\tilde{g}}e^{-2\phi}(\tilde{\cR}
+4(\nabla\phi)^2-F^2)\;.
}
For the solution~\estring{}, the string metric is (after rescaling
$r \rightarrow {\sqrt{2} Q \over \gamma} r$)
\eqn\esmet{
d\tilde{s}^2= - {r^2 \over \gamma^4} \left( 1 - {2 \sqrt{2} \gamma^2
M \over Q r} \right) dt^2 + {1 \over \left( 1 - {2 \sqrt{2} \gamma^2
M \over Q r} \right)} dr^2 + r^2 d\Omega^2 \; .
}
This string metric shares many of the same properties with the
Einstein metric~\emetb. The event horizon is located at $r_h= {2
\sqrt{2} \gamma^2 M \over Q}$.  Again, the only curvature singularity
is located at $r=0$ and there is no extremal limit.  However, the
string coupling $e^{2\phi} \propto r$ is blowing up for large $r$, so
string loop corrections should be important there.

Now consider the magnetically charged version of~\estring{} in string
metric.  The magnetically charged solution can be obtained by taking
$\phi \rightarrow - \phi$ and leaving the Einstein metric
unchanged. In the string metric, the magnetically charged solution is
\eqn\esmetb{
d\tilde{s}^2 = - {2 Q^2 \over \gamma^4} \left(1 - {4 M \over r}\right)
dt^2 + {2 Q^2 \over r^2 \left(1 - {4 M \over r}\right)} dr^2 + 2 Q^2
d\Omega^2 \; .
}
The metric~\esmetb\ is a well-known solution. The angular part of the
metric has become a cylinder with radius $\sqrt{2} Q$. The metric is
equivalent to taking the extremal limit of the magnetically charged
black hole of~\ghs\ in coordinates that retain the event horizon. It
is also equivalent to the direct product of the two-dimensional black
hole~\witten\ and a cylinder.

\subsec{General Maxwell coupling}

It is straightforward to generalize the solution~\estring{} to
arbitrary coupling $a$. Using the ansatz~\ehansatz\ in~\eeomb{}, we
find the solution
\eqna\egena
$$\eqalignno{
U(r) = & {r^{{2 \over 1 + a^2}} \over \gamma^2} \left(1 - {2 (1 + a^2) M
\over a^2 r}\right) \; , & \egena{a} \cr
N = & { a^2 \over 1 + a^2} \; , & \egena{b} \cr
\phi(r) & = - {1 \over 2 a} \log \left( {Q^2 (1 + a^2) \over \gamma^2}
\right) + {a \over 1 + a^2} \log r \; . & \egena{c}
}$$
where $M$ is the quasilocal mass~\emquas.  It is easy to check
that~\egena{} reduces to~\estring{} when $a=1$.  If
$a^2\rightarrow\infty$, the metric reduces to the Schwarzschild black
hole. Again, \egena{} has no smooth transition to the $Q=0$ case.  It
has an event horizon at
\eqn\egenah{
r_h = {2 M (1 + a^2) \over a^2} \; ,
}
and a singularity at $r=0$. The scalar curvature for~\egena{} is
\eqn\escala{
\cR = {2 a^2 r^{{-2 a^2 \over 1 + a^2}} \over (1 + a^2)^2 \gamma^2}
\left(1 - {2 (1 + a^2) M \over a^2 r} \right) \; ,
}
which again vanishes for large $r$.  We believe that they are the
first examples of black hole solutions in four-dimensional general
relativity or string gravity which have charge without having an
extremal limit.

The solution~\egena{} is a special case of the general solutions to
the equations of motion~\eeomb{} which have appeared in~\horhora\ (for
$a=1$) and~\polwil. Both these works overlooked the special solution
that has a regular horizon but is not asymptotically flat. For
example, setting $n=4$ and $A=1$ in eqs.~(B8-10) of~\polwil\ leads
to~\egena{}. When $A \ne 1$, the solution of~\polwil\ is
asymptotically flat, and that is the solution considered in~\polwil.

The behavior of~\egena{} for large $r$ depends on the value of
$a$. Consider a radial null geodesic parameterized by $\lambda$. Along
the geodesic, we have $0 = - U(r) \dot{t}^2 + \dot{r}^2/U(r)$, where
$\dot{x}$ represents differentiation with respect to $\lambda$. Thus
$t = \pm \int 1/U(r) dr + {\rm const}$ along the geodesic. If $t
\rightarrow \infty$ as $r \rightarrow \infty$, then it takes infinite
affine parameter to reach $r = \infty$. This is the case for $a^2 \ge
1$. For this case, the causal structure is like the Schwarzschild
causal structure. If $t \rightarrow {\rm const}$ as $r \rightarrow
\infty$, then it takes finite affine parameter to reach $r = \infty$,
and we must extend our coordinates beyond there. This happens when
$a^2 < 1$. Though it is impossible to integrate $1/U(r)$ for general
$a$, special values such as $a^2 = {1 \over 3}$ do yield to
analysis. They have a causal structure similar to anti-de Sitter
space.

For a metric of the form~\emeta, the temperature at a horizon $r_h$ is
given by
\eqn\etemp{
T = {1 \over 4 \pi} U'(r_h) \; .
}
For the solution~\egena{}, the temperature is
\eqn\etempa{
T = {1 \over 4 \pi \gamma^2} r_h^{{1 - a^2 \over 1 + a^2}} \; .
}
Thus, the black hole temperature decreases as the mass decreases for
$a^2 < 1$, is independent of mass for $a^2=1$, and increases as the
mass decreases for $a^2 > 1$.

\newsec{Exact solutions for $V(\phi)=2\Lambda e^{2b\phi}$}

In this section, we consider the action \esdef\ with a Liouville-type
potential
\eqn\eliouv{
V(\phi)=2\Lambda e^{2b\phi} \; .
}
We are unable to solve~\eeomb{} in general with this $V$, but using
the ansatz~\ehansatz\ in~\eeomb{} gives solutions for special values
of $\Lambda$ and $b$.  In particular, three types of exact
electrically charged solutions exist.

\medskip
\noindent{(i)}  Using the notation of~\emeta, \ehansatz, and
\ephiansatz, the first solution is
\eqna\etypea
$$\eqalignno{
U_1(r) = & r^{{2 a^2 \over 1 + a^2}}
\left({ 1 + a^2 \over (1 - a^2) \gamma^2}
- {2 (1 + a^2) M \over \gamma^{2} r}
  + { Q^2 (1 + a^2) e^{2 a \phi_0}\over \gamma^{4}r^2} \right) \; , &
\etypea{a}
\cr
N = & {1 \over 1 + a^2} \; , & \etypea{b} \cr
\phi_1 = & - {a \over 1 + a^2 } \; , & \etypea{c} \cr
\Lambda = & -{ a^2 \over \gamma^2 (1 - a^2)} e^{- {2
\phi_0 \over a }} \; , & \etypea{d} \cr
b = & {1\over a} \; . & \etypea{e} \cr
}$$
Note that this solution does not exist for the string case where $a
=1$, and the limit $a \rightarrow 0$ gives the standard
Reissner-Nordstrom solution. The parameter $M$ in~\etypea{a} is the
quasilocal mass from~\emquas. As before, the only singularity is at
$r=0$.

The solution~\etypea{} is qualitatively somewhat different
than~\egena{}. Eq.~\etypea{} has a well-defined $Q=0$ limit, and has
horizons at
\eqn\ehora{
r_\pm = (1 - a^2) M \left(1 \pm \sqrt{1 - {Q^2 e^{2 a \phi_0} \over
\gamma^2 (1 - a^2) M^2}} \right) \; .
}
If $a^2 < 1$, then~\etypea{} has two horizons, and an extremal limit
when
\eqn\eexta{
Q^2_{\rm ext} e^{2 a \phi_0} =
 (1 - a^2) \gamma^2  M^2 \; .
}
If $Q^2 > Q^2_{\rm ext}$, then no solution exists with a smooth
horizon.  Using~\etemp, the temperature at the outer horizon is
\eqn\etypeat{
T = {(1 + a^2) M \over 2 \pi \gamma^2 r_+^{{3 + a^2 \over 1 + a^2}}}
\left( r_+ - {Q^2 e^{2 a \phi_0} \over \gamma^2 M} \right) \; ,
}
which vanishes in the extremal limit.  On the other hand, if $a^2 >
1$, then~\ehora\ has just one positive root, which exists for any
$Q^2$.

The causal structures of the black holes in \etypea{} is difficult to
obtain even in the uncharged case since it is generally impossible to
calculate $\int 1/U(r) dr$ for general $a$. When $a=0$, the causal
structure is the Reissner-Nordstrom one.  As another example, take the
$a^2= {1\over 2}$, $Q=0$ case. One can then perform the integration
and see that the causal structure is exactly the same as the
Schwarzschild one. When $0 < Q^2 < Q^2_{\rm ext}$, it can be deduced
that the causal structure is the Reissner-Nordstrom one.  Causal
structures of other values of $a$ within the range $0 < a^2 <1$ can be
constructed similarily, and are qualitatively the same as the
Reissner-Nordstrom solution.  For $a^2 > 1$, the solution \etypea{}
has a naked singularity at $r=0$ and a cosmological horizon at the
single horizon $r_h$, beyond which $U_1$ is negative and $r$ behaves
like a time coordinate.

At first glance, the chargeless action~\esdef\ with a
potential~\eliouv, is identical to the dimensionally reduced action
of~\refs{\kimcho,\wilt}, which has a internal space with $l$
dimensions. In~\wilt, it was shown that there is no positive mass
Liouville black hole solution with a dilaton of the
form~\ephiansatz. The apparent contradiction can be resolved by noting
that the black hole solutions in~\etypea{} correspond to $l<0$, a
negative number of internal dimensions.

\medskip
\noindent{(ii)} The second solution is
\eqna\etypeb
$$\eqalignno{
U_2(r) = & r^{{2 \over 1 + a^2}} \left( {1
+ a^2 \over (1 - a^2) \gamma^2}
\left( -1 + {2 Q^2 e^{{2 a \phi_0}} \over\gamma^{2}}
\right)
- {2 (1 + a^2) M \over a^2 \gamma^{2} r} \right) &
\etypeb{a} \cr
N = & {a^2 \over 1 + a^2} \; , & \etypeb{b} \cr
\phi_1 = & {a \over 1 + a^2} \; , & \etypeb{c} \cr
\Lambda = & { e^{2 a \phi_0} \over \left( 1 - a^2 \right)
\gamma^2} - { (1+ a^2) Q^2 e^{4 a \phi_0} \over \left( 1- a^2\right)
\gamma^{4}} \; , & \etypeb{d} \cr
b = & -  a \; . & \etypeb{e} \cr
}$$
The $a=1$ limit of \etypeb{} again is ill defined.  Setting $\Lambda =
0$ in \etypeb{d} and solving for $\phi_0$ recovers~\egena{}. The
solution~\etypeb{} has a single event horizon at
\eqn\ehorb{
r_h  = {2 M (1 - a^2) \over a^2 \left( -1 + {2 Q^2 e^{{2 a \phi_0}} \over
\gamma^{2}}
\right)} \; .
}
Regular horizons only exist when the right-hand-side of~\ehorb\ is
positive. Thus the extremal limits of~\etypeb{} correspond to $r_h
\rightarrow 0$ or $r_h \rightarrow \infty$. When a horizon exists, the
causal structure of~\etypeb{} depends on $a$ in the same manner as the
$V=0$ case.  The temperature of \etypeb{} at $r_h$ is
\eqn\etempb{
T = { 1+ a^2 \over 4 \pi (1 - a^2) \gamma^2}
\left( -1 + {2 Q^2 e^{{2 a \phi_0}} \over
\gamma^{2}} \right) r_h^{{1 - a^2 \over 1 + a^2}}
}

\medskip
\noindent{(iii)} The third solution is
\eqna\etypec
$$\eqalignno{
U_3(r) = & r^{2 \over 1 + a^2} \left( {1 \over \gamma^2} - {2 (1 + a^2)
M \over a^2 \gamma^2 r} +
{(1 + a^2)^{{2 a^2 + 1 \over a^2}} \Lambda
Q^{2 \over a^2} \over (1 - 3 a^2) a^2 \gamma^{2 \over a^2}}
r^{-{2 (1 - a^2) \over 1 + a^2}} \right) & \etypec{a} \cr
N = & { a^2 \over 1 + a^2} \; , & \etypec{b} \cr
\phi_1 = & { a \over \left( 1 + a^2 \right)} \; ,
& \etypec{c} \cr
\phi_0 = & - {1 \over 2 a} \log\left({Q^2 (1 + a^2) \over \gamma^{2}}
\right) \; , & \etypec{d} \cr
b = & - {1\over a} \; . & \etypec{e} \cr
}$$
We have chosen $M$ in \etypec{} to be the mass~\emquas\ when $\Lambda
= 0$.  The solution~\etypec{} is closely related to~\egena{}, as can
be seen by setting $\Lambda = 0$ in \etypec{}.  Because of~\etypec{d},
the solution~\etypec{} only exists for nonzero $Q$. Unlike the
previous two solutions, \etypec{} does exist for $a = 1$.
Notice that
if $a^2 > 1$, then $U_3 \rightarrow -\Lambda \times \infty$ as
$r\rightarrow \infty$. If $a^2 < 1$, then $U_3
\rightarrow + \infty$ as $r\rightarrow \infty$.

The spacetimes associated with the solutions \etypec{} exhibit a
variety of possible causal structures depending on the values of $a$
and $\Lambda$.  We can obtain the causal structure for all possible
values of $a$ and $\Lambda$ by finding the roots of $U_3(r)$, noting
that this problem reduces to finding the intersection of a curve
linear in $r$ with a variable power of $r$.  Unfortunately, because of
the nature of the exponents of $r$ in~\etypec{}, it is not possible to
find explicitly the locations of the horizons where $U_3(r_h) = 0$ for
arbitrary $a$.

First consider $a=1$, where a single event horizon exists at
\eqn\etypech{
r_h = {4 M \over 1 - 4 Q^2 \Lambda} \; .
}
{}From~\etypech, we see that a regular horizon only exists for $4 Q^2
\Lambda < 1$. At this horizon, the temperature is
\eqn\etypect{
T = {1 \over 4\pi\gamma^2} (1 - 4 Q^2 \Lambda) \; ,
}
which approaches zero in the extremal limit $4 Q^2 \Lambda = 1$, and
is independent of $M$.

When $\Lambda > 0$ and $a^2 < {1 \over 3}$ or $a^2 > 1$, there can be
zero, one, or two horizons, depending on the relative magnitudes of
$M$ and $\Lambda$.  One horizon occurs at the extremal limit when
\eqn\etypecl{
\Lambda = {a^2 \gamma^{{2 (1 - a^2) \over a^2}} \over (1 +
a^2)^{{1 + a^2 \over a^2}} Q^{{2 \over a^2}}} r_{\rm ext}^{2 {1 -
a^2 \over 1 + a^2}} \equiv \Lambda_{\rm ext}\; ,
}
where
\eqn\etypecx{
r_{\rm ext} = {(1 - 3 a^2)(1+a^2) \over a^2(1 - a^2)}{M} \; .
}
is the location of the extremal horizon.  If $\Lambda > \Lambda_{\rm
ext}$, then no horizons exist. Finally, if $\Lambda < 0$ or ${1 \over
3} < a^2 < 1$, then~\etypec{} has a single horizon.  The possible
cases are listed in table~1, with the equality signs denoting an
extremal limit.

{\midinsert
\centerline{{\bf Table~1:} The possible horizons for eq.~\etypec{} }
\medskip
\centerline{\vbox{\offinterlineskip
\hrule
\halign{&\vrule#&\strut\quad\hfil#\quad\cr
height2pt&\omit&&\omit&&\omit&\cr
&Solution \etypec{} &&$\Lambda<0$\hfil && $\Lambda>0$\hfil &\cr
height2pt&\omit&&\omit&&\omit&\cr
\noalign{\hrule}
height2pt&\omit&&\omit&&\omit&\cr
& && && (\C,\ \O) $\quad \Lambda <\Lambda_{\rm ext}$&\cr
&$1 < a^2$&& \O\hfil && (\C=\O) $\quad \Lambda =\Lambda_{\rm ext}$&\cr
& && && \B  $\qquad \Lambda >\Lambda_{\rm ext}$&\cr
height2pt&\omit&&\omit&&\omit&\cr
\noalign{\hrule}
height2pt&\omit&&\omit&&\omit&\cr
&${1\over 3} < a^2 < 1$&& \O\hfil && \O\hfil &\cr
height2pt&\omit&&\omit&&\omit&\cr
\noalign{\hrule}
height2pt&\omit&&\omit&&\omit&\cr
& && &&(\O,\ \I) $\quad \Lambda >\Lambda_{\rm ext}$ &\cr
&$a^2 < {1\over 3}$ && \O\hfil&& (\O=\I) $\quad \Lambda =\Lambda_{\rm
ext}$\hfil &\cr
& && && \N  $\qquad \Lambda <\Lambda_{\rm ext}$ &\cr
height2pt&\omit&&\omit&&\omit&\cr}
\hrule}}
\smallskip
{\settabs
\+{\it Notation:}
& {\C}--Cosmological horizon\quad\quad\quad
& {\O}--Outer black hole horizon
\cr
\+{\it Notation:}
& {\O}--Outer black hole horizon
&{\I}--Inner black hole horizon
\cr
\+
&{\B}--Cosmological singularity
&{\N}--Naked singularity
\hfill\cr\+
& {\C}--Cosmological horizon
&\cr
}
\endinsert}

\newsec{Exact solutions for
$V(\phi)=2\Lambda_1e^{2b_1\phi}+2\Lambda_2e^{2b_2\phi}$}

The final type of potential that we will study is the sum of two
Liouville type terms:
\eqn\evtwo{
V(\phi) = 2 \Lambda_1 e^{2 b_1 \phi} + 2 \Lambda_2 e^{2 b_2 \phi} \; .
}
This generalizes further the potential~\eliouv. If $b_1 = b_2$,
then~\evtwo\ reduces to~\eliouv, so we will not repeat these
solutions. Requiring $b_1 \ne b_2$ again gives three classes of
solutions.

\medskip
\noindent{(a)} The first solution is identical to the
solution~\etypea{}, except for the modifications
\eqna\etypeta
$$\eqalignno{
U(r) = &
U_1(r) - { \Lambda_2 (1 + a^2)^2 e^{2 a \phi_0}
\over (3 - a^2)} r^{2 \over 1 + a^2} \; , & \etypeta{a}
\cr
b_2 = & a  \; . & \etypeta{b}
}$$
Notice that for $a=1$ only, $b_1 = b_2$, which is the previous
solution~\etypea{}.

To analyze the causal structures associated with this case, we can
proceed as before, noting that the roots of $U(r)$ may be found by
finding the intersection of a quadratic with a variable power.
The results are given in table~2.

{\midinsert
\centerline{{\bf Table~2:} The possible horizons for eq.~\etypeta{}}
\medskip
\centerline{\vbox{\offinterlineskip
\hrule
\halign{&\vrule#&\strut\quad\hfil#\quad\cr
height2pt&\omit&&\omit&&\omit&\cr
&Solution \etypeta{} &&${\Lambda}_2<0$\hfil && ${\Lambda}_2>0$\hfil &\cr
height2pt&\omit&&\omit&&\omit&\cr
\noalign{\hrule}
height2pt&\omit&&\omit&&\omit&\cr
&$3 < a^2$&& \C\hfil && \C\hfil &\cr
height2pt&\omit&&\omit&&\omit&\cr
\noalign{\hrule}
height2pt&\omit&&\omit&&\omit&\cr
& &&(\C,\ \O,\ \I)\hfil&& &\cr
&$1< a^2 < 3$ && (\C,\ \O=\I) or (\C=\O,\ \I) && \C\hfil &\cr
& && \C\ or (\C=\O=\I)\hfil&& &\cr
height2pt&\omit&&\omit&&\omit&\cr
\noalign{\hrule}
height2pt&\omit&&\omit&&\omit&\cr
& &&(\O,\ \I)\hfil &&(\C,\ \O,\ \I)\hfil &\cr
&$a^2 < 1$ && (\O=\I)\hfil&& (\C,\ \O=\I) or (\C=\O,\ \I) \hfil &\cr
& && \N\hfil&& \C\ \  or (\C=\O=\I)\hfil &\cr
height2pt&\omit&&\omit&&\omit&\cr}
\hrule}}
\endinsert}

We see from table~2 that there is now the possiblity of having
spacetimes with three horizons, whose causal structure resembles that
of Reissner-Nordstrom-de Sitter spacetime.  The extremal cases in
table~2 can be approached in a variety of ways, depending upon the
relative magnitudes of $M$, $Q$ and $\Lambda_2$. We leave the
derivation of the extremal values of these quantities as an exercise
for the reader.

The general form of the temperature at the outer event horizon
for these solutions is
\eqn\etempfa{
T = {r_O^{-{3+a^2 \over 1+a^2}}\over{2\pi\gamma^2}}
\left(r_O^2 - M(a^2-3)r_O - 2{Q^2 e^{2a\phi_0}\over \gamma^2}\right) \;
}
where $r_O$ denotes the location of the outer horizon when it exists.
In general $r_O$ must be solved for numerically, although closed form
solutions can be obtained for special values of $a$.

\medskip
\noindent{(b)}
The second solution is identical to the solution~\etypeb{}, except
for the modifications
\eqna\etypetb
$$\eqalignno{
U(r) = & U_2(r) + { \Lambda_2 (1 + a^2)^2 e^{- { 2\phi_0 \over a }} \over a^2
(1 - 3 a^2)}
r^{{2 a^2 \over 1 + a^2}} \; , & \etypetb{a} \cr
b_2 = & - {1 \over a } \; . & \etypetb{b}
}$$
Again, when $a=1$, $b_1 = b_2$. For arbitrary $a$, the extra power of
$r$ in~\etypetb{a} makes it impossible to find explicitly the
positions of the horizons, but we can determine the number and type of
horizons. The possible causal structures are given in table~3, where
we have used $\hat{Q} \equiv {1 + a^2 \over (1 - a^2) \gamma^2}
\left({2 Q^2 \over \gamma^2} e^{2 a \phi_0} - 1 \right)$.  The
extremal cases in table~3 occur for certain values $M=M_{\rm ext}$ of
the mass parameter; the explicit derivation is similar to the previous
cases and we shall not reproduce it here.

The temperature of the outer horizon in this case is given by
\eqn\etempfb{
T = {r_O^{-2{a^2 \over 1+a^2}}\over{2\pi\gamma^2}}\left[
\left({2 Q^2 e^{{2 a \phi_0}} \over\gamma^{2}} - 1 \right)r_O
- {3a^2-1 \over a^2}M \right]
}
where again $r_O$ denotes the location of the outer horizon (when it
exists) and must be solved for numerically except for special values
of $a$ and/or $Q$.

{\midinsert
\centerline{{\bf Table~3:} The possible horizons for eq.~\etypetb{}}
\medskip
\centerline{\vbox{\offinterlineskip
\def\tablerule{\noalign{\hrule}}
\halign{\strut#&\vrule#\tabskip=1em plus2em&\hfil#& \vrule#& \hfil#\hfil&
\vrule#&\hfil#\hfil& \vrule#&\hfil#\hfil& \vrule#& \hfil#&
\vrule#\tabskip=0pt\cr\tablerule
\omit&height2pt&\omit&&\multispan3 &&\multispan2 &&\cr
&&\hfil&& \multispan3\hfil$\quad{\Lambda}_2<0$\hfil
&&\multispan2\hfill$\quad{\Lambda}_2>0$\hfil && \cr\tablerule
\omit&height2pt&\omit &&\omit &&\omit &&\omit &&\omit &\cr
&&\omit Solution \etypetb{}  && \omit $\hat{Q}<0$&&
\omit $\hat{Q}>0$&&
\omit $\hat{Q}<0$&&
\omit$\hat{Q}>0$&\cr
\omit&height2pt&\omit &&\omit &&\omit &&\omit &&\omit &\cr
\tablerule
\omit&height2pt&\omit &&\omit &&\omit &&\omit &&\omit &\cr
&& && && && &&(\C,\ \O)\hfil  &\cr
&&$1 < a^2$&&\O\hfil&& \O\hfil && \B\hfil &&(\C=\O)\hfil  &\cr
&& && && && &&\B\hfil\hfil &\cr
\omit&height2pt&\omit &&\omit &&\omit &&\omit &&\omit &\cr
\tablerule
\omit&height2pt&\omit &&\omit &&\omit &&\omit &&\omit &\cr
&& &&(\C,\ \O)\hfil && && && &\cr
&&${1\over 3}< a^2 < 1$&&(\C=\O)\hfil && \O\hfil&&\N\hfil && \O\hfil &\cr
&& && \B\hfil\hfil && && &&   &\cr
\omit&height2pt&\omit &&\omit &&\omit &&\omit &&\omit & \cr
\tablerule
\omit&height2pt&\omit &&\omit &&\omit &&\omit &&\omit & \cr
&& && && && &&(\O,\ \I)\hfil &\cr
&&$a^2 < {1\over 3}$&&\N\hfil&&\O\hfil &&\C\hfil &&(\O=\I)\hfil &\cr
&& && && && &&\N\hfil\hfil &\cr
\omit&height2pt&\omit &&\omit &&\omit &&\omit &&\omit & \cr
\tablerule}}}
\endinsert}

\medskip
\noindent{(c)}
Finally, a third class of solutions exists that has no counterpart in
the solutions with $V$ of the form~\eliouv. These solutions exist for
$0 < N < 1$ with $N \ne {1 \over 2}$, and satisfy
\vfil\eject
\eqna\etypetc
$$\eqalignno{
U(r) = & {1 \over (2N - 1) \gamma^2} r^{2(1 - N)}
- r_0 r^{1 - 2 N}   & \etypetc{a} \cr
& +
{2 Q^2 e^{{2 a \phi_0}} r^{2 \left( 1 - 2 N \pm 2 a \sqrt{N (1 -
N)}\right)}\over
\left(1- 2 N \pm 2 a \sqrt{N (1 - N)}\right) \left(1 - 3 N \pm a \sqrt{N (1
- N)}\right) \gamma^{4}}  \; ,\cr
b_1 = & \mp {N \over \sqrt{N(1 - N)}} \; , & \etypetc{b}
\cr
\Lambda_1 = & {(N - 1) e^{\pm{2 N \phi_0 \over
\sqrt{N (1 - N)}}} \over (2 N - 1) \gamma^2} \; , & \etypetc{c} \cr
b_2 = & a \mp { 2 N \over \sqrt{N (1 - N)}} \; , &
\etypetc{d} \cr
\Lambda_2 = & - {Q^2 \left(1 - N \pm a \sqrt{N (1 - N)}\right) \over \left(1 -
3 N \pm a \sqrt{N(1 - N)}\right) \gamma^{4}} e^{\pm {4 N \phi_0
\over \sqrt{N(1 - N)}}} \; , & \etypetc{e} \cr
\phi_1 = & \pm \sqrt{N(1 - N)} \; , & \etypetc{f}
}$$
where $r_0$ is a constant related to the mass.

The possible causal structures for this case depend upon the relative
values of $a$, $b_1$, $r_0$ and $\check{Q}\equiv 2 Q^2 e^{2 a
\phi_0}(1- 2 N \pm 2 a \sqrt{N (1 - N)})^{-1} (1 - 3 N \pm a \sqrt{N
(1- N)})^{-1} \gamma^{-4}$. The derivation is similar to the previous
cases and the results are given in table~4.  Despite the appearance of
the additional parameter $N$, there are at most two horizons for any
solution in this class.  We leave the derivation of the temperature
and extremal cases as an exercise.

{\midinsert
\centerline{{\bf Table~4:} The possible horizons for eq.~\etypetc{}}
\medskip
\centerline{\vbox{\offinterlineskip
\def\tablerule{\noalign{\hrule}}
\halign{\strut#&\vrule#\tabskip=1em plus2em&\hfil#& \vrule#& \hfil#\hfil&
\vrule#&\hfil#\hfil& \vrule#&\hfil#\hfil& \vrule#& \hfil#&
\vrule#\tabskip=0pt\cr\tablerule
\omit&height2pt&\omit&&\multispan3 &&\multispan2 &&\cr
&&\hfil&& \multispan3\hfil$\quad{\check{Q}}<0$\hfil
&&\multispan2\hfill$\qquad{\check{Q}}>0$\hfil && \cr\tablerule
\omit&height2pt&\omit &&\omit &&\omit &&\omit &&\omit &\cr
&&Solution \etypetc{}  && \omit $b_1<1$&&
\omit $b_1>1$&&
\omit $b_1<1$&&
\omit$b_1>1$&\cr
\omit&height2pt&\omit &&\omit &&\omit &&\omit &&\omit &\cr
\tablerule
\omit&height2pt&\omit &&\omit &&\omit &&\omit &&\omit &\cr
&& && &&(\C,\ \O)\hfil && &&  &\cr
&&$2ab_1>1+b_1^2$&&\O\hfil&&(\C=\O)\hfil && \B\hfil && \O\hfil &\cr
&& && &&\B\hfil\hfil && &&  &\cr
\omit&height2pt&\omit &&\omit &&\omit &&\omit &&\omit &\cr
\tablerule
\omit&height2pt&\omit &&\omit &&\omit &&\omit &&\omit &\cr
&& &&(\C,\ \O)\hfil  && &&&& &\cr
&&$1+b_1^2>2ab_1>0$&&(\C=\O)\hfil&&\O\hfil &&\B\hfil && \O\hfil &\cr
&& &&\B\hfil && && &&   &\cr
\omit&height2pt&\omit &&\omit &&\omit &&\omit &&\omit & \cr
\tablerule
\omit&height2pt&\omit &&\omit &&\omit &&\omit &&\omit & \cr
&& && && && &&(\O,\ \I)\hfil &\cr
&&$0>ab_1$&&\C\hfil&&\O\hfil &&\B\hfil &&(\O=\I)\hfil &\cr
&& && && && &&\N\hfil\hfil &\cr
\omit&height2pt&\omit &&\omit &&\omit &&\omit &&\omit & \cr
\tablerule}}}
\endinsert}

\newsec{Solutions in $n$ dimensions}

The solutions given in the previous sections are valid in $n=4$
dimensions, but can be generalized to dimensions $n \ge 4$. Using the
action~\eddef, we can derive the equations of motion
(following~\polwil)
\eqna\edeom
$$\eqalignno{
\cR_{ab} = & {4 \over n - 2} \left( \partial_a \phi \partial_b \phi +
{1 \over 4}
g_{ab} V \right) + 2 e^{ -{4 a \phi \over n - 2}} \left( F_{ac}
{F_b}^c - {1 \over 2 (n - 2)} g_{ab} F^{2} \right) \;, &
\edeom{a} \cr
0 = & \partial_a \left[ \sqrt{-g} e^{- {4 a \phi \over n - 2}} F^{ab}
\right] \;, & \edeom{b} \cr
\nabla^2\phi = & {n - 2 \over 8} {\partial V \over \partial
\phi} - {a \over 2} e^{-{4 a \phi \over n - 2}} F_{ab} F^{ab} \; . &
\edeom{c} \cr
}$$
We can again choose a coordinate system satisfying \emeta.  The
Maxwell field can be chosen to be an isolated electric charge
satisfying \edeom{b}
\eqn\edf{
F_{tr} = e^{{4 a \phi \over n - 2}} {Q \over R^{n - 2}} \; ,
}
which generalizes \efsol. With the coordinate system \emeta\ and the
Maxwell field \edf, the equations of motion \edeom{} reduce to~\polwil
\eqna\edafter
$$\eqalignno{
{1 \over R^{n - 2}} {d \over dr}\left[ R^{n - 2} U {d \phi \over dr}
\right] = & {n - 2 \over 8} {d V \over d\phi} + a e^{{4 a \phi \over n
-2}} {Q^2 \over R^{2 (n - 2)}} \; , & \edafter{a} \cr
{1 \over R} {d^2 R \over dr^2} = & - {4 \over (n - 2)^2} \left( {d
\phi \over dr} \right)^2 \; , & \edafter{b} \cr
{1 \over R^{n - 2}} {d \over dr} \left[ U {d \over dr} \left( R^{n -
2} \right) \right] = & (n - 2) (n - 3) {1 \over R^2} - V - 2 e^{{4 a
\phi \over n - 2}} {Q^2 \over R^{2(n - 2)}} \; . & \edafter{c} \cr
}
$$

\bigskip
\subsec{Solutions with $V = 0$ for arbitrary $n$}

First consider solutions of~\edafter{} where the potential vanishes,
$V(\phi) = 0$. Making the ansatz~\ehansatz\ $R(r) = \gamma r^N$, we
find a solution of \edafter{} with
\eqna\edsola
$$\eqalignno{
U(r) = & { \left( (n - 3)^2 + a^2 \right)^2 \over \gamma^2 \left( n -
3 + a^2 \right)^2} r^{{ 2 (n - 3)^2 \over (n - 3)^2 + a^2}} - { 4
\left( (n - 3)^2 + a^2 \right) \over a^2 (n - 2) \gamma^{n - 2}} M
r^{{(n - 3) (n - 3 - a^2) \over (n - 3)^2 + a^2}} \; , & \edsola{a}
\cr
N = & { a^2 \over (n - 3)^2 + a^2} \; ,& \edsola{b} \cr
\phi(r) = & - {n - 2 \over 4 a} \log\left( {2 Q^2 (n - 3 + a^2) \over
(n - 2) (n - 3)^2 \gamma^{2(n - 3)}} \right) + {a (n - 2) (n - 3)
\over 2  \left( (n - 3)^2 + a^2 \right)} \log r \; , & \edsola{c}
\cr}$$
where $M$ is the mass as defined by~\emquas. The metric is independent
of $Q$, but because of the $\phi$ dependence on $Q$, solutions only
exist for nonzero values of $Q$.  The solution~\edsola{} has an event
horizon at
\eqn\edhor{
r_h = \left( {4 M (n - 3 + a^2)^2 \over a^2 (n - 2) ((n - 3)^2 + a^2)
\gamma^{n - 4} } \right)^{{(n - 3)^2 + a^2 \over (n - 3) (n - 3 +
a^2)}} \; .
}
The temperature is therefore
\eqn\edtemp{
T = { (n - 3)((n - 3)^2 + a^2) \over 4 \pi \gamma^2 (n - 3 + a^2)}
r_h^{{(n - 3)^2 - a^2 \over (n - 3)^2 + a^2}} \; .
}
Temperature decreases as mass decreases for $a^2 < (n - 3)^2$, is
independent of mass for $a^2 = (n - 3)^2$, and increases as mass
decreases for $a^2 > (n - 3)^2$.  The causal structure for the
solution~\edsola{} is similar the solution~\egena{} when $n=4$, except
that the boundary between Schwarzschild-like causal structure and
anti-de Sitter-like causal structure is at $a^2 = (n - 3)^2$.

\subsec{Solutions with $V(\phi) = 2 \Lambda e^{2 b \phi}$}

Solutions to \edafter{} also exist for nonzero $V$. As in four
dimensions, the simplest case is
\eqn\evdef{
V(\phi) = 2 \Lambda e^{2 b \phi} \; .
}
Using the ansatz~\ehansatz, solutions exist only for certain values of
$\Lambda$ and $b$. There are now three types of solutions
to~\edafter{}.

\medskip
\noindent{(i)} The first solution has the form
\eqna\edtypea
$$\eqalignno{
U_1(r) = & r^{{2 a^2 \over 1 + a^2}}
\Bigg({ (1 + a^2)^2 (n -3) \over (1 - a^2) \gamma^2  (n - 3 + a^2)}
- {4 (1 + a^2) M \over (n - 2) \gamma^{n - 2}}
     r^{-{n - 3 + a^2 \over 1 + a^2}} \cr
& + {2 Q^2 (1 + a^2)^2 \over (n - 2) (n - 3 + a^2) \gamma^{2(n - 2)}} e^{{4 a
\phi_0 \over n - 2}} r^{-{2 (n - 3 + a^2) \over 1 + a^2}} \Bigg) \; , &
\edtypea{a}
\cr
N = & {1 \over 1 + a^2} \; , & \edtypea{b} \cr
\phi_1 = & - {a (n - 2) \over 2 (1 + a^2)} \; , & \edtypea{c} \cr
\Lambda = & -{ (n - 2) (n - 3) a^2 \over 2 \gamma^2 (1 - a^2)} e^{- {4
\phi_0 \over a (n - 2)}} \; , & \edtypea{d} \cr
b = & {2 \over a (n - 2)} \; . & \edtypea{e} \cr
}$$
Note that this solution does not exist for the string case where $a
=1$. The solution~\edtypea{} has horizons given by
\eqn\edhora{
r_\pm^{{n - 3 + a^2 \over 1 + a^2}} = { 2 (1 - a^2) (n - 3 + a^2) M
\over (1 + a^2) (n - 2) (n - 3) \gamma^{n - 4}} \left( 1 \pm \sqrt{ 1
- {  (1 + a^2)^2 (n - 2) (n - 3) Q^2 e^{{4 a \phi_0 \over n - 2}}
\over 2 (1 - a^2) (n - 3 + a^2)^2 \gamma^2 M^2}} \right)
}
Thus, as in the $n=4$ case~\etypea{}, when $a^2 < 1$, \edtypea{} has
two horizons, and an extremal limit when
\eqn\edexta{
Q^2_{\rm ext} e^{{4 a \phi_0 \over n - 2}} =
 {2 (1 - a^2) (n - 3 + a^2)^2 \gamma^2  \over (n - 2) (n - 3) (1 +
a^2)^2 } M^2 \; .
}
Again, the temperature at the outer horizon is
\eqn\edtypeat{
T = {(n - 3 + a^2) M \over \pi (n - 2) \gamma^{n - 2} r_+^{{2 n - 5 +
a^2 \over 1 + a^2}}} \left( r_+^{{n - 3 + a^2 \over 1 + a^2}} - {(1 +
a^2) Q^2 e^{{4 a \phi_0 \over n - 2}} \over (n - 3 + a^2) \gamma^{n -
2} M}\right) \; ,
}
which vanishes in the extremal limit.  If $a^2 > 1$, then~\edtypea{}
has a single horizon.

\medskip
\noindent{(ii)} The second solution has the form
\eqna\edtypeb
$$\eqalignno{
U_2(r) = & r^{{2 (n - 3)^2 \over (n - 3)^2 + a^2}} \Bigg( {((n - 3)^2
+ a^2)^2 \over (n - 3 + a^2) ((n - 3)^2 - a^2) \gamma^2} \left( -1 +
{2 Q^2 e^{{4 a \phi_0 \over n - 2}} \over (n - 3) \gamma^{2(n - 3)}}
\right) \cr
& - {4 ((n - 3)^2 + a^2) M \over a^2 (n - 2) \gamma^{n - 2}}
r^{- { (n - 3) (n - 3 + a^2) \over (n - 3)^2 + a^2}} \Bigg) &
\edtypeb{a} \cr
N = & {a^2 \over (n - 3)^2 + a^2} \; , & \edtypeb{b} \cr
\phi_1 = & {a (n - 2) (n - 3) \over 2 \left( (n - 3)^2 + a^2 \right)}
\; , & \edtypeb{c} \cr
\Lambda = & { (n - 2) (n - 3)^3 \over 2 \left( (n - 3)^2 - a^2 \right)
\gamma^2} e^{{4 a \phi_0 \over (n - 2) (n - 3)}} - {(n - 3) (n - 3 +
a^2) Q^2 \over \left( (n - 3)^2 - a^2\right) \gamma^{2(n - 2)}}
e^{{4 a \phi_0 \over n - 3}} \; , & \edtypeb{d} \cr
b = & - {2 a \over (n - 2) (n - 3)} \; . & \edtypeb{e} \cr
}$$
Setting $\Lambda = 0$ in \edtypeb{d} and solving for $\phi_0$ recovers
\edsola{}. The solution~\edtypeb{} has a single event horizon at
\eqn\edhorb{
r_h^{{(n - 3)(n - 3 + a^2) \over (n - 3)^2 + a^2}} = {4 M (n - 3 +
a^2) ((n - 3)^2 - a^2) \over a^2 (n - 2) ((n - 3)^2 + a^2) \gamma^{n -
4} \left( -1 +
{2 Q^2 e^{{4 a \phi_0 \over n - 2}} \over (n - 3) \gamma^{2(n - 3)}}
\right)} \; .
}
Regular event horizons only exist when the right-hand-side of~\edhorb\
is positive. Thus the extremal limits of~\edtypeb{} correspond to $r_h
= 0$ or $r_h = \infty$. The temperature is
\eqn\edtempb{
T = { (n - 3) ((n - 3)^2 + a^2) \over 4 \pi ((n - 3)^2 - a^2) \gamma^2}
\left( -1 + {2 Q^2 e^{{4 a \phi_0 \over n - 2}} \over (n - 3)
\gamma^{2(n - 3)}} \right) r_h^{{(n - 3)^2 - a^2 \over (n - 3)^2 +
a^2}}
}

\medskip
\noindent{(iii)} The third solution has the form
\eqna\edtypec
$$\eqalignno{
U_3(r) = & { \left( (n - 3)^2 + a^2 \right)^2 \over \gamma^2 (n - 3 +
a^2)^2} r^{{ 2 (n - 3)^2 \over (n - 3)^2 + a^2}} - {4 ((n - 3)^2 +
a^2) M \over a^2 (n - 2) \gamma^2} r^{{ (n - 3) (n
- 3 - a^2) \over (n - 3)^2 + a^2}} & \edtypec{a} \cr
& + {2 \Lambda Q^{{2 (n -3) \over a^2}} 2^{{n - 3 \over a^2}} \left(
(n - 3)^2 + a^2 \right)^2 (n - 3 + a^2)^{{n - 3 \over a^2}} \over a^2
\gamma^{{2 (n- 3)^2 \over a^2}} \left( (n- 3)^2 - (n - 1) a^2 \right)
(n - 2)^{{n - 3 + a^2 \over a^2}} (n - 3)^{{2 (n- 3) \over a^2}}}
r^{{2 a^2 \over (n - 3)^2 + a^2}} \; ,  \cr
N = & { a^2 \over (n - 3)^2 + a^2} \; , & \edtypec{b} \cr
\phi_1 = & { a (n- 2)(n-3) \over 2 \left( (n - 3)^2 + a^2 \right)} \; ,
& \edtypec{c} \cr
\phi_0 = & - {n -2 \over 4 a} \log\left({2 Q^2 (n- 3 + a^2) \over (n -
2) (n - 3)^2 \gamma^{2(n - 3)}} \right) \; , & \edtypec{d} \cr
b = & - {2 (n - 3) \over (n -2) a} \; . & \edtypec{e} \cr
}$$
Setting $\Lambda = 0$ in \edtypec{} easily reduces
to~\edsola{}. Because of~\edtypec{d}, the solution~\edtypec{} only
exists for nonzero $Q$. Unfortunately, because of the nature of the
exponents of $r$ in~\edtypec{}, it is not possible to find explicitly
the locations of the horizons where $U_3(r_h) = 0$.

\subsec{Solutions with $V(\phi) = 2 \Lambda_1 e^{2 b_1 \phi} + 2
\Lambda_2 e^{2 b_2 \phi}$}

Solutions also exist for
\eqn\evdtwo{
V(\phi) = 2 \Lambda_1 e^{2 b_1 \phi} + 2 \Lambda_2 e^{2 b_2 \phi} \; .
}
Restricting to $b_1 \ne b_2$ reveals three classes of solutions.

\medskip
\noindent{(a)} The first solution is identical to the
solution~\edtypea{}, except for the modifications
\eqna\edtypeta
$$\eqalignno{
U(r) = & U_1(r) - { 2 \Lambda_2 (1 + a^2)^2 e^{4 a \phi_0 \over n - 2}
\over (n - 2) (n - 1 - a^2)} r^{2 \over 1 + a^2} \; , & \edtypeta{a}
\cr
b_2 = & {2 a \over n - 2} \; . & \edtypeta{b}
}$$

\medskip
\noindent{(b)}
The second solution is identical to the solution~\edtypeb{}, except
for the modifications
\eqna\edtypetb
$$\eqalignno{
U(r) = & U_2(r) + { 2 \Lambda_2 ((n - 3)^2 + a^2)^2 e^{- { 4 (n - 3)
\phi_0 \over a (n - 2)}} \over a^2 (n - 2) ((n - 3)^2 - (n - 1) a^2)}
r^{{2 a^2 \over (n - 3)^2 + a^2}} \; , & \edtypetb{a} \cr
b_2 = & - {2 (n - 3) \over a (n - 2)} \; . & \edtypetb{b}
}$$

\medskip
\noindent{(c)}
Finally, a third class of solutions exists that has no counterpart in
the solutions with $V$ of the form~\evdef. These solutions exists for
$0 < N < 1$ with $N \ne {1 \over 2}$, and satisfy
\eqna\edtypetc
$$\eqalignno{
U(r) = & {n - 3 \over (2N - 1) (1 - 4 N + n N) \gamma^2} r^{2(1 - N)}
- r_0 r^{1 - N(n - 2)}   & \edtypetc{a} \cr
& + {2 Q^2 e^{{4 a \phi_0 \over n - 2}} r^{2 \left( 1 - N (n - 2) \pm 2 a
\sqrt{N (1 -
N)}\right)}\over \left(1
- N (n - 2) \pm 2 a \sqrt{N (1 - N)}\right) \left(1 - N (n - 1) \pm a \sqrt{N
(1
- N)}\right) \gamma^{2(n -2)}}  \; ,\cr
b_1 = & \mp {2 N \over (n - 2) \sqrt{N(1 - N)}} \; , & \edtypetc{b}
\cr
\Lambda_1 = & {(N - 1)(n - 2)(n - 3) e^{\pm{4 N \phi_0 \over (n - 2)
\sqrt{N (1 - N)}}} \over 2 (2 N - 1) \gamma^2} \; , & \edtypetc{c} \cr
b_2 = & {2 a \over n - 2} \mp { 2 N \over \sqrt{N (1 - N)}} \; , &
\edtypetc{d} \cr
\Lambda_2 = & - {Q^2 \left(1 - N \pm a \sqrt{N (1 - N)}\right) \over \left(1 -
N (n
- 1) \pm a \sqrt{N(1 - N)}\right) \gamma^{2(n - 2)}} e^{\pm {4 N \phi_0
\over \sqrt{N(1 - N)}}} \; , & \edtypetc{e} \cr
\phi_1 = & \pm {1 \over 2} (n - 2) \sqrt{N(1 - N)} \; , & \edtypetc{f}
}$$
where $r_0$ is a constant.

\newsec{Conclusions}

To sum up, we have obtained several one-parameter families of
non-asymptotically flat ((anti-) de Sitter) electrically charged $SSS$
black hole solutions for the action \esdef\ with a zero potential,
Liouville potential and a two-term exponential potential. In the
$V(\phi)=0$ case, this family of charged black holes (which includes
the string theoretic case) has two unique properties: (i) they do not
have inner horizons, and (ii) $Q$ can take any nonzero value without
causing the event horizon to vanish or become singular.  The former
property can be found in the black holes in~\gibmae\ but the second
property has no previous analogue in general relativity or string gravity. The
magnetically charged string black hole has an additional property: it
is a product of the $2D$ black hole with a constant two sphere, in
terms of the string metric. For the Liouville black holes, we derived
three families of solutions. For the first family~\etypea{}, the
solutions have outer and inner horizons if the extremal limit~\eexta\
between $Q^2$ and $M^2$ is not violated. It has no well-defined $a=1$
limit.  For the second family~\etypeb{}, the solution has only one
event horizon.  Again, the limit $a=1$ is not well-defined. For the
third family~\etypec{}, which has a well-defined $a=1$ limit, the
solutions have possible outer, inner, or cosmological horizons.  Three
different families of $SSS$ solutions are also obtained in the case of
a two-term exponential potential. The first two families~\etypeta{}
and~\etypetb{} are the modifications of~\etypea{} and~\etypeb{}
respectively.  The third family has no counterpart to the solutions for
the single potential~\eliouv.  The conditions of existence of outer,
inner and cosmological horizons for the three families of solutions
are summarized in tables 2, 3 and 4 respectively.  The Hawking
temperature of all the above black hole solutions are also computed.
Finally, we generalized all the above solutions to $n$ dimensions.

We close by commenting on further possible extensions of our work. It
is worthwhile to mention that the black hole solutions~\egena{} can be
interpreted as black $p$-branes in $(4+p)$ dimensions which are
solutions to the $(4+p)$-dimensional Einstein-Maxwell (or with a
$p$-form) action, provided that $a=\sqrt{{p\over p+2}}$ (the string
case $a=1$ implies a diverging $p$)~\townsend. Using these solutions
as examples, it would be interesting to see whether curvature
singularities in the four dimensional solutions can be completely resolved or
become much milder, by viewing the solutions as reductions of
higher-dimensional objects, with or without serious modification in
higher-dimensional gravitational actions or metrics.

Note that asymptotically flat rotating charged black hole solutions
were discussed and investigated in~\horhorc\ for the action \esdef\
with $V(\phi)=0$. No explicit solutions have yet been found except for
$a=\sqrt{3}$~\fzb\ and for $a = 1$ when the string three-form
$H_{abc}$ is included~\sen. It would be interesting to add angular
momentum to the string black hole solutions we obtained in section~3
and~4 to see if it is possible to extend the family of solutions we
obtained to the rotating case.  Generalization of all the solutions
derived in this paper to arbitrary $n\geq 4$ with both electric and
magnetic charge present is another problem of interest.  One can also
attempt to construct any asymptotically flat or (anti-) de Sitter
black holes for the two-term exponential potential.  We note that a
partially closed-form chargeless black hole solution for
action~\esdef\ with a more elaborate potential than~\evtwo\ has been
obtained in~\bechlech\ with an exponentially decaying dilaton.
Finally, we note that using the results in $2+1$ dilaton gravity
in~\chaman, Maki and Shiraishi derived several static solutions to
Einstein equations coupled with antisymmetric tensor fields in
$(2+l+1)$ dimensions. The solutions describe a product of a
$(2+1)$-dimensional circularly symmetric spacetime and an
$l$-dimensional internal maximally symmetric manifold, and they can be
applied to various supergravity models~\makshib. One may consider a
similar work in $(3+l+1)$ dimensions by using the results in the
present paper. We intend to relate further details elsewhere.

\bigskip
\centerline{\bf Acknowledgements}

This work was supported in part by the Natural Science and Engineering
Research Council of Canada. RBM would like to thank DAMTP for its
hospitality.

\listrefs
\bye